Combined breakdown of a dielectric nanolayer to form a Josephson bridge

A.V. Krevsun, S.I. Bondarenko, V.P. Koverya

B. Verkin Institute for Low Temperature Physics and Engineering of the National Academy of Sciences of Ukraine, Kharkiv 61103, Ukraine
E-mail: bondarenko@ilt.kharkov.ua

Electrical breakdown of a dielectric nanolayer between film electrodes under the combined action of direct current and capacitor discharge current makes it possible to form Josephson bridges with a reproducible resistance exceeding 1 Ω. A new feature of the formation of a bridge during electrical breakdown, which is preceded by a series of preliminary breakdowns (auto breakdowns) of a dielectric nanolayer, has been discovered. The mechanism of the process and the role of the thickness of the cathode film in the formation of the bridge are discussed.

1. Introduction

It is known [1-7] that as a result of electrical breakdown of a thin dielectric layer between two metal electrodes, a conducting channel is formed in it in the form of a metal bridge. Almost all numerous studies of the formation of such bridges were previously carried out using non-superconducting massive electrodes. For the first time, studies of some properties of bridges made of superconducting materials formed as a result of electrical breakdown were carried out in 1970-1976 in B.I. Verkin Institute for Low temperature Physics and Engineering (ILTPE) [8, 9, 10]. At the same time, the use of film electrodes was also new. The purpose of these studies was to elucidate the possibility of using electrical breakdown to create a superconducting weak link between films with Josephson properties. In particular, three methods were tested to achieve the breakdown of a dielectric: a smooth increase in the electric voltage across it from a direct voltage source, a discharge of an electric capacitor through it, and a single rectangular voltage pulse [10]. The material of the bridges formed by the electrical breakdown was lead or a lead-tin compound, from which a thin cathode film was made. The features of the cathode metal film transfer through the dielectric during breakdown are an important and little studied issue. The result of the initial studies of the formation of superconducting bridges during the electrical breakdown of a dielectric was an experimental proof of the fundamental possibility of creating such Josephson weak links [9, 10]. In the future, based on these film bridges, it was planned to develop miniature film superconducting direct current (DC) quantum interferometer with high sensitivity and stability of parameters even under significant external mechanical influences.

At the same time, a number of questions concerning the patterns of bridge formation remained unexplained. The answer to some of them was obtained by us in a recent work [11], in which an electrical breakdown of the dielectric oxide layer of a niobium film with a thickness of 30 nm was carried out using a 50% indium–50% tin alloy film with a thickness of 100 nm as a cathode. In particular, for the first time it was found that with a smooth (less than 1 V/s) increase in voltage up to breakdown, the minimal and reproducible total value of the bridge resistance ($R_\Sigma \approx 3\ \Omega$), consisting of the intrinsic resistance of the bridge $R_b$ and the resistance of the contact conductor, was achieved at a breakdown current of 10 mA. In this case, the indicated resistance values at a temperature of $T=300K$ were reproducible on dozens of bridges. The diameter (about 25 nm) of the bridge itself with a resistance of $R_b \approx 1\ \Omega$ was determined by calculation. It was experimentally proved that the bridge material is identical to the indium-tin alloy of a cathode film. At the same time, the manufacture of Josephson bridges (JB) by the electrical breakdown method is much simpler than by known traditional methods.

It was also shown that in the temperature range of 300-77K, the dependence of the resistance of the cathode film and bridges on temperature with its decrease is linearly decreasing, and their temperature coefficients of resistance (TCR) are within $(2.0 \pm 0.3) \times 10^{-3}$ K$^{-1}$. Note that the TCR of massive indium and tin are (4.9 and 4.4) $\times 10^{-3}$ K$^{-1}$, respectively.

The importance of determining the resistance value of the bridge at T=300K is that it correlates with the value of its normal resistance (NR) near the superconducting transition and the critical current of the JB in the superconducting state, which determines its Josephson properties [12].

At present, there are a number of unresolved issues in the technology of manufacturing JB using breakdown. These include questions about the possibility to form a reproducible JB with $R_b > 1$ Ohm, about influence of the thickness of the cathode film on the formation of JB, as well as the use of alternative methods of electrical breakdown of the dielectric and their influence on the NR value of the resulting JB. In particular, such a method can be the breakdown of niobium oxide using the combined electrical action on the oxide of direct current and the current of a capacitor connected in parallel and charged to the breakdown voltage.

The purpose of this work was to find answers to these questions through experimental studies.

2. Setting up the experiment.
Figure 1 shows the electrical circuit of the device for the combined breakdown of niobium oxide in order to form JB.

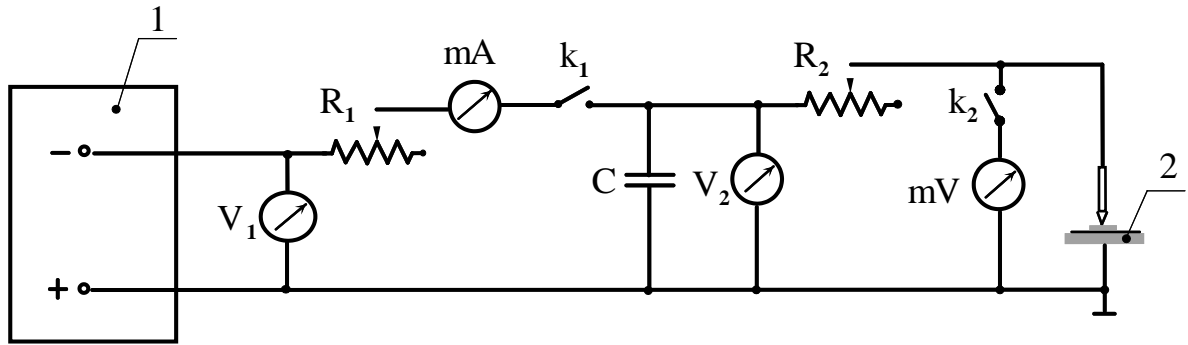

Fig.1. Electrical diagram of the device for the combined breakdown of niobium oxide.
1 – direct voltage source, 2 – structure of InSn and Nb films separated by a nanolayer of niobium oxide, $C$– capacitor; $k_1$, $k_2$ – keys; $R_1$, $R_2$– variable resistors; mA is a miliammeter; $V_1$, $V_2$, mV – voltmeters and millivoltmeter.

The design of the film structure (2), consisting of an alloy film of indium (In) - tin (Sn) with a thickness of 100 nm as a cathode and a niobium (Nb) film of the same thickness as an anode, separated by a layer of niobium oxide with a thickness of 30 nm, similar to the structure studied by us in [11]. It also describes a contact device for applying voltage to the films and a method for measuring the voltage on it during electrical breakdown of the oxide. First, its value is set on the voltage source, slightly exceeding the breakdown voltage of the oxide. After that, the key $K_1$ is closed. The constant voltage source (1) through the limiting resistor $R_1$ provides an increase in the current through the oxide and charges the capacitor $C$. When the breakdown voltage of the oxide (2) is reached, the capacitor is discharged to the breakdown oxide through the discharge resistance $R_2$. Resistor $R_1$ limits the charge current of capacitor $C$, as well as the direct current through the resistance formed after the breakdown of the bridge. The resistance $R_1$ in the course of experiments is always much greater than the resistance $R_\Sigma$ of the bridge and the discharge resistance. Resistor $R_2$ limits the amount of discharge current of the capacitor $C$ during electrical breakdown and practically does not affect the amount of direct current through the oxide. Based on the measured direct current through the JB and the voltage on it after the breakdown, according to Ohm's law, it is possible to determine the resistance $R_\Sigma$ of the bridge. In addition, the surface of the cathode film was visually inspected after breakdown using an optical microscope. The variable elements of the device circuit (Fig. 1) were the limiting resistance $R_1$, capacitance $C$ and discharge resistance $R_2$, the values of which changed respectively: for $R_1$– within $1 \div 10$ k$\Omega$, for $R_2$– within $0 \div 50$ $\Omega$, capacitance $C$– within $40 \div 180$ microfarads.

The breakdown of the dielectric was carried out in the following way. With the key $K_1$ open, the voltage of the power source (1) was set at 15 V and did not change during the experiment. After closing this switch, there was a gradual increase in the voltage across the capacitor and oxide, as well as the current from the direct voltage source through the oxide (with time constant $\tau = R_1C$). Upon reaching the breakdown voltage (12–13 V), breakdown of the oxide occurred, which was accompanied by a sharp drop in the voltage across it. After the switch $K_2$ was closed, measurements were made of the direct current through the formed JB and the residual voltage across it. The value of the measuring current through the JB was usually set equal to 10 mA using a direct voltage source and limiting resistance $R_1$.

As our experiments have shown, the combined method (CM) of the breakdown of niobium oxide, compared to the method of smooth (about 1 V/s) increase in voltage (MSIV) on the oxide without using a capacitor [11] it allows not only to form reproducible bridges with a minimum resistance ($R_b \approx 1\ \Omega$, $R_\Sigma \approx 3\ \Omega$), but also increase the resistance of the bridges with a reproducible value. In addition, CM made it possible to detect the unusual phenomenon of repeated breakdowns (auto-breakdowns) of the oxide with some parameters of the circuit (Fig. 1) of the device for their implementation. When using CM, the minimum total bridge resistance ($R_\Sigma \approx 3\ \Omega$), consisting of the minimum resistance of the bridge itself $R_b \approx 1\ \Omega$ and the resistance of the voltage supply conductors ($R_c = 2\ \Omega$), is formed when using a capacitor with a capacitance of 180 μF and zero discharge resistance. At the same time, in contrast to the MSIV [11], the resistance $R_\Sigma$ practically does not depend on the change in the limiting resistance in the range $R_1 = 1 \div 10$ kΩ. An increase in the discharge resistance $R_2$ above zero leads to the formation of bridges with large and reproducible resistance values $R_\Sigma$ and $R_b$ (Fig.2).

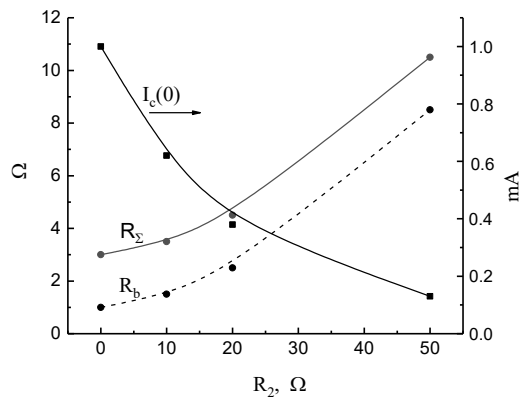

Fig.2. Dependences of the total resistance ($R_\Sigma$) of the bridges, the bridges themselves ($R_b$) and their critical current ($I_c$) on the value of the discharge resistance $R_2$ in the capacitor circuit.

Changing the capacitance of the capacitor from 180 µF to 40 µF at fixed values of both limiting (1 kΩ) and discharge resistances (10 Ω) does not lead to noticeable changes in the total resistance of the bridges (about 3.5 Ω). Figure 2 also shows the corresponding resistance values ($R_b$) of the JB itself and their critical currents at a temperature of $T = 0$ K. In this case, the values of critical currents were obtained by calculation using formula (6) from our previous article [11].

Based on the existing theory [3, 11] of the resistance of bridges formed by electrical breakdown of a thin insulator layer, it can be assumed that the obtained increase of resistances ($R_b > 1$ Ω) of reproducible JBs at their fixed length ($s = 30$ nm) is caused by a decrease in their diameters ($a < 25$ nm) according to formula (3) in [11]. With resistances in the range $R_b = 1 \div 8.5$ Ω, obtained in this work, this corresponds (while maintaining the purity of the bridge material) to a decrease in the JB diameter from 25 nm to 14 nm. As can be seen in Fig. 2, the critical current of the JB at $T = 0$K decreases from 1 mA to 0.13 mA.

The cathode film, depending on the parameters of the circuit (Fig. 1), can be either homogeneous or inhomogeneous after breakdown. The surface of the cathode film in the form of a disk is homogeneous and contains only a single darkening at the breakdown site, if $C = 180$ µF, $R_1 = 1$ kΩ, $R_2 = 20-50$ Ω (Fig. 3a). The film remains homogeneous even at $R_2 = 10$ Ω, if $R_1 = 10$ kΩ with the same capacitor capacitance (Fig. 3b). If the discharge resistance is in the range of $0 \div 10$ Ω, then with a limiting resistance of 1 kΩ and a capacitance value of the capacitor in the range of $40 \div 180$ µF, the film surface becomes inhomogeneous after breakdown (Fig. 4a,b).

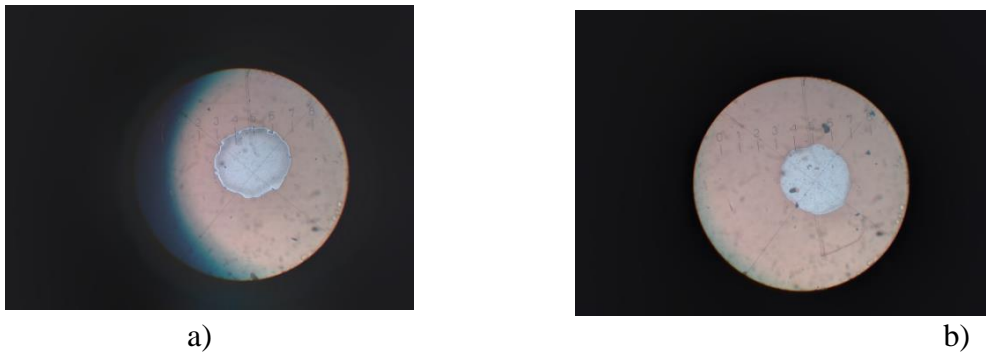

a)  b)

Fig. 3. Appearance of a homogeneous cathode film (a bright spot with a diameter of about 0.5 mm) on niobium oxide after its combined breakdown in a circuit with a 180 µF capacitor, a) with a discharge resistance of 20 Ω and a limiting resistance of 1 kΩ, b) with a discharge resistance of 10 Ω and a limiting resistance of 10 kΩ.

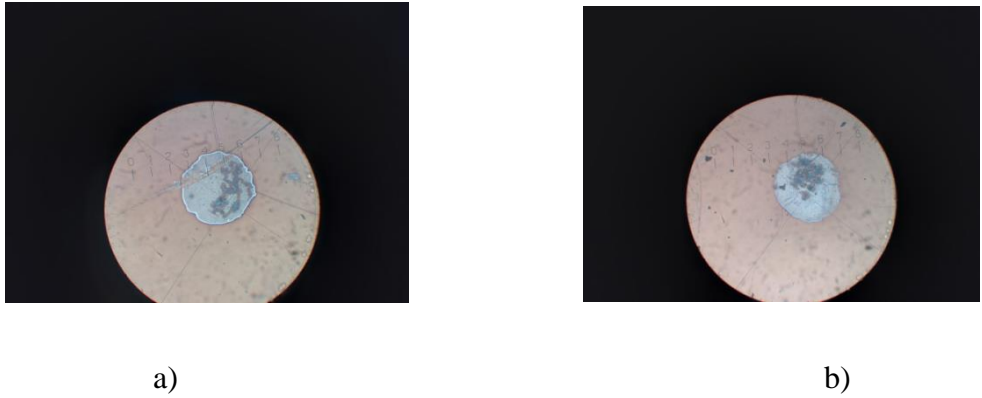

a)                                                                                  b)

Fig. 4. Appearance of an inhomogeneous cathode film (a bright spot with a diameter of about 0.5 mm) on niobium oxide after its combined breakdown in a circuit with a 180 µF capacitor and a limiting resistance of 1 kΩ, a) with a discharge resistance of 0 Ω, b) with a discharge resistance 10 Ω.

Let us proceed to the discussion of experimental observations. Let's start by finding out the reason for the appearance of an inhomogeneous cathode film as a result of breakdown. The minimum bridge resistance, as noted earlier, is formed at zero discharge resistance in the capacitor circuit and is accompanied by the appearance of film inhomogeneity. It should be noted that when using a cathode in the form of a film disk with a diameter of 0.5 mm, breakdown of the oxide under it can occur at the weakest point in terms of electrical strength within the disk diameter. We do not know the location of this point in advance. The path of the breakdown current lies from the contact pressure electrode on the cathode film along its section between it and the breakdown point. Just before the breakdown of the oxide, it is heated by a weak current from a direct voltage source [3] and a capacitor. This creates conditions for the breakdown of the dielectric. It can be assumed that at the moment of breakdown (for a time of about 100 µs [3]) a bridge is formed with a minimum resistance and then a capacitor with a capacitance $C = 180$ µF is discharged onto this resistance. Estimates show that the average density of the pulsed current (during the time $\tau = R_\Sigma C \approx 500$ µs) from the capacitor through the film section not far from the bridge can be about $3 \times 10^6$ A/cm². At such a high current density, its pulsed heating can occur with a large increase in temperature, oxidation in air, melting and evaporation of the metal. For example, it is known that fuses in electrical appliances blow out in air at a current density of $10^5$ A/cm². The strong heating and oxidation of the film at the breakdown site is evidenced by the tint colors of its inhomogeneous sections. As a result of the evaporation of a section of the film, a break occurs in the current path from the contact electrode to the bridge. At the same time, as a result of heating the niobium oxide adjacent to the cathode film, its permittivity increases and its breakdown voltage decreases [13, 14], which leads to the appearance of conditions for the next breakdown. There are other current paths from the contact electrode along the film to a new breakdown point and the formation of a bridge in this oxide site weakened by heating. Such a process can be repeated in a short time (this time is planned to be measured in future studies) several times until the voltage on the capacitor, which is periodically shunted by emerging bridges, decreases to a value below the breakdown value, and only one highly conductive bridge remains between the electrodes. As a result, the cathode film is covered with traces of high-density current paths, which form the observed pattern of its inhomogeneity.

The possibility of repeated breakdowns (auto breakdowns) is also confirmed by comparing the capacitor energy with the evaporation energy of the film section adjacent to the breakdown site. Let us estimate the amount of energy, which is necessary for the evaporation of such a section of an indium-tin alloy film with a thickness of 100 nm and a diameter of about 0.1 mm, considering the heating process to be adiabatic. If we assume that the average heat of vaporization of the indium-tin alloy is about $2.6 \times 10^5$ J/mol, then the evaporation of a film with a mass of about $10^{-8}$ g requires an energy input equal to $1.3 \times 10^{-5}$ J. The value of the energy of the capacitor with a capacitance of 180 μF at a breakdown voltage of 13 V is about $2 \times 10^{-2}$ J, which significantly exceeds the energy required to form a series of such bridges which can be called autobridges.

It follows from our experience that if the discharge current density through the resulting bridge is limited by increasing the discharge resistance (for example, to 20 or 50 Ω), then the formed primary bridge and the uniformity of the cathode film surface are preserved, since the energy supplied to the film decreases and there are no conditions for repeated breakdowns.

As already noted, the cathode film also remains homogeneous after breakdown by a 180 uF capacitor with a discharge resistance of 10 Ω, and a limiting resistance of 10 kΩ (Fig. 3b), although it is inhomogeneous with a limiting resistance of 1 kΩ (Fig. 4b). Such an effect of the limiting resistance can be explained by the fact that the sum of direct current and pulsed current with a limiting resistance of 10 kΩ is less than with a limiting resistance of 1 kΩ. As a result, the current density through the cathode film decreases and the film is not destroyed.

It should also be noted that it is possible to achieve a decrease in the current density through the cathode film by increasing its thickness and, thereby, reducing its overheating. It is planned to experimentally investigate this possibility in our future works.

It should be noted that bridges with increased resistances (1.5÷8.5 Ω) at a temperature of 300K, obtained by the combined method, are well reproducible from sample to sample. Therefore, this method can be used to obtain more wide range of the required critical currents of the JBs, since the critical current of a superconducting JB is inversely proportional to its normal resistance $R_b$ ($T_c$) at the superconducting transition temperature $T_c$ [12]. This resistance, in turn, is determined by the resistance value $R_b$ (300) at a temperature of 300K and the positive temperature coefficient α of resistance (TCR) determined by us for an indium-tin alloy film ($2 \times 10^{-3}$ K$^{-1}$): $R_b$ ($T_c$) = $R_b$ (300) [ 1- α $T_c$]..

Finally, we note one more aspect of the influence of the cathode film thickness on the parameters of the bridge formed by the breakdown. With its too small thickness, the microscopic volume of the film melted by the discharge current and located above the place of breakdown of the oxide may not be enough to form a metal bridge between the film electrodes during the liquid stage of its existence. To avoid this situation, it is necessary that the thickness of the cathode film be greater than the thickness of the oxide.

Conclusion.

The electrical breakdown of a niobium oxide nanolayer with a thickness of 30 nm between two film electrodes (the cathode is an indium-tin alloy film, the anode is a niobium film) under the combined action of a constant voltage source and a charged capacitor was possible to discover new features of the formation of Josephson film nanobridges.

Firstly, this is the possibility of forming nanobridges with a reproducible resistance, not only equal to the minimum value ($R_b \approx 1$ Ω), typical for the method of smoothly increasing the voltage on the oxide, but also with high resistances (1.5÷8.5 Ω). In our experiments, this is achieved by using a discharge resistance in the range from 10 to 50 Ω in a capacitor circuit with a capacity of 40 to 180 microfarads. In this case, the calculated diameter of the bridge and its critical current at a temperature of $T = 0K$ decrease, respectively, from 25 nm to 14 nm and from 1 mA to 0.13 mA.

Secondly, this is a violation of the homogeneity of the cathode film in the process of repeated auto breakdowns of the oxide at certain parameters of the breakdown device circuit. The observed phenomenon can be explained by the significant amplitude of the discharge current density (about $3 \times 10^6$ A/cm$^2$) of the capacitor, at which it significantly exceeds the allowable value for a thin (100 nm) cathode film. As a result, at the moment of breakdown, it is destroyed in the area of its contact with the bridge. This automatically increases the probability of the next breakdown of the oxide at another point of the cathode film disk above the oxide with a reduced value of the breakdown voltage. Within a short time, several such auto breakdowns occur until the voltage on the capacitor decreases to a value less than the breakdown value, and one highly conductive bridge appears between the electrodes. It follows from the studies performed that in order to avoid repeated breakdowns (auto breakdowns) and save the film cathode, it is necessary to have such parameters of the breakdown scheme and such a film thickness at which the discharge current density through the film does not exceed the critical value.

The work was carried out with the financial support of the Leading Program of the National Academy of Sciences of Ukraine "Fundamental research on the most important problems of natural sciences" (section "Quantum nano-sized superconducting systems: theory, experiment, practical implementation"). State registration number of the work is 0122U001503.

References.